# Stimulated Raman-induced Beam Focusing

## MINHAENG CHO[1,2*]

[1]Center for Molecular Spectroscopy and Dynamics, Institute for Basic Science (IBS), Seoul 02841, Republic of Korea.
[2]Department of Chemistry, Korea University, Seoul 02841, Republic of Korea.
* mcho@korea.ac.kr

**Abstract:** Stimulated Raman scattering, employing a pump and a Stokes beam, exhibits itself through both the Raman loss observed in the pump beam and the Raman gain in the Stokes beam. This phenomenon finds application in spectroscopy for chemical analyses and microscopy for label-free bioimaging studies. Recent efforts have been made to implement super-resolution Raman microscopy using a doughnut-shaped pump, Stokes, or depletion beam. In this study, it is shown that the amplitude and phase of the pump or Stokes beam undergo significant modulation through the stimulated Raman process when they are configured as one of the higher-order Laguerre-Gauss modes, achieved using appropriate spiral phase plates or spatial light modulators. The resulting intensity distributions of the pump and Stokes beams are determined by a superposition of multiple Laguerre-Gauss modes that are coupled through nonlinear Raman gain and loss processes. Calculation results are used to elucidate the limitations associated with super-resolution coherent Raman imaging with a toroidal pump or Stokes beam. This stands in contrast with the stimulated emission depletion fluorescence microscopy technique, which lacks a fundamental limit in the spatial resolution enhancement.

## 1. Introduction

Spontaneous Raman scattering is an inelastic process that entails the one-photon annihilation of incident light. This process excites a single molecular vibration, simultaneously resulting in the creation of a low-frequency photon. Stimulated Raman scattering [1] differs from spontaneous Raman scattering by introducing another light called Stokes beam to stimulate the same Raman scattering process. While the theoretical and experimental exploration of stimulated Raman scattering (SRS) with high-intensity lights is extensive, it has gained significant attention for its applications in bioimaging over the past two decades [2-4]. SRS microscopy offers a distinct advantage over other optical imaging methods that necessitate site-specific labels, such as fluorescent dyes, proteins, or sometimes bulky nanoparticles. Unlike fluorescence-based microscopy techniques, SRS or coherent anti-Stokes Raman scattering [5, 6] (CARS) microscopy can be used to obtain spatial distribution and functional information about the major chemical components of biological systems in a label-free manner [3, 7-11].

Spontaneous Raman microscopy, employing a focused pump beam, has been utilized for detecting vibrational signatures of target molecules in cells or tissues without the need for fluorescent labels [12-14]. However, the well-known issue of a weak spontaneous Raman scattering cross-section has impeded the widespread adoption of spontaneous Raman microscopy as a practical imaging tool. In this regard, CARS microscopy has attracted significant attention due to its strong signal, typically several orders of magnitude larger than the spontaneous Raman process [15]. Nevertheless, CARS spectra or microscopic images inevitably suffer from a non-resonant electronic background signal, limiting its applicability in high-resolution bioimaging. As a response, SRS microscopy, employing a lock-in amplification scheme, has emerged as a popular vibrational imaging technique. SRS microscopy is free from non-resonant background interference and features a well-defined point spread function (PSF), which makes it an attractive choice for high-resolution bioimaging applications.

Nevertheless, the spatial resolution of SRS microscopy remains confined by Abbe's diffraction limit. In super-resolution fluorescence microscopy, diverse approaches have effectively overcome this limitation. The stimulated emission depletion (STED) method, in theory, imposes no spatial resolution limit [16, 17]. The width of the PSF in an STED image can be significantly reduced by increasing the ratio of doughnut-shaped STED beam intensity to saturation intensity. Similarly, in coherent Raman scattering microscopy, noteworthy strategies, both theoretically proposed and experimentally demonstrated, include the utilization of a spiral phase plate to generate a doughnut-shaped beam at the focal region to achieve super-resolution Raman imaging beyond the diffraction limit [18-23].

In this context, Frontiera and coworkers [18] conducted interesting experiments using a scheme, hereafter referred to as the Frontiera scheme, which incorporates a Gaussian pump-1 beam ($\omega_p$), a doughnut depletion ($\omega_d = \omega_p$) or pump-2 beam, and a Gaussian Stokes beam ($\omega_S$) (see Figure 1(a) for this Frontiera approach). They proposed that the stimulated Raman process, involving the Gaussian pump-1 and Stokes beams, generates vibrational coherence of molecules in the region where the pump-1 and Stokes beams overlap. Subsequently, the toroidal depletion (pump-2) beam, with a frequency identical to that of the Gaussian pump-1 beam, is employed to deplete the vibrational coherence. However, the propagation of the Stokes beam, nonlinearly interacting with the strong doughnut-shaped depletion (pump-2) beam through the stimulated Raman gain (SRG) process, was not thoroughly investigated.

An alternative approach proposed in Ref. [19] employed a combination of a weak Gaussian pump beam and a strong doughnut-shaped Stoke-1 beam (Figure 1(b)). In the overlapping region of these two beams, the stimulated Raman loss (SRL) of the pump beam could induce a narrowing of the pump beam. This narrowed pump beam is then used to initiate another SRS process with a third (Stokes-2) beam. This alternative technique, referred to as the Cho scheme in this paper, has been suggested as a potentially effective method to overcome the diffraction limit of SRS microscopy. However, it has not been thoroughly investigated, considering the stimulated Raman coupling of the Gaussian pump and doughnut Stokes beams, and addressing and solving the wave equations of the two beams derived from Maxwell's equations. Here, it is worth noting that Gong and Wang [20] theoretically proposed an interesting idea for breaking the diffraction limit by using a pair of the Gaussian pump and doughnut-shaped Stokes beams.

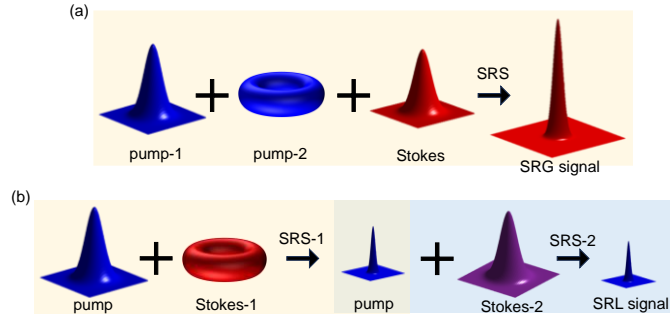

**Figure 1.** Schematic illustrations of the Frontiera (a) and Cho (b) schemes. (a) In the Frontiera scheme, two-beam SRS involves pump-1 and Stokes beams. Pump-2, with the same frequency as pump-1, is used to deplete the vibrational coherence (Ref. [18]) in the molecule where the doughnut pump-2 beam overlaps with the Gaussiaan Stokes beam. Thus, the Stokes beam resulting from the SRS processes by undepleted molecules becomes narrower than the width under the diffraction limit. (b) In the Cho scheme [19], two SRS processes occur simultaneously and compete with each other for a single pump beam. The first SRS-1 results in an intensity loss of the pump through the SRL of the pump induced by an intense doughnut-shaped Stokes beam. Then, the focused (narrowed) pump interacts with the Gaussian Stokes-2 beam through the SRS-2 process.

They suggested that the doughnut Stokes beam can be employed to saturate SRS at the rim of a focused Gaussian pump beam. However, this approach requires a powerful pump and Stokes beam, potentially leading to photodamage to biosamples under coherent Raman imaging studies.

In this study, the pump and Stokes beams are expanded in terms of radial Laguerre-Gauss (LG) modes, and the resulting coupled equations for the mode expansion coefficients are solved. It is shown that in the Cho scheme, the stimulated Raman process modulates the phase and amplitude of the paraxial Gaussian pump beam through interaction with the doughnut-shaped Stokes beam. Similarly, in the Frontiera scheme, the Gaussian Stokes beam becomes mixed with higher-order radial LG modes in the presence of a strong doughnut-shaped pump beam (pump-2 in Figure 1(a)). The theoretical and numerical results from calculations are used to elucidate the limitations of these two approaches in enhancing the spatial resolution of coherent Raman microscopy.

## 2. Stimulated Raman scattering: Radial Laguerre-Gauss mode expansion

### 2.1. Coupled wave equations for the pump and Stokes beams

From Maxwell's equation, the following wave equation for $\vec{E}$ with nonlinear polarization $\vec{P}^{(3)}$, a source term,[24-27] is considered:

$$\nabla^2 \vec{E} - \frac{n^2}{c^2}\frac{\partial^2}{\partial t^2}\vec{E} = \frac{4\pi}{c^2}\frac{\partial^2}{\partial t^2}\vec{P}^{(3)},\tag{1}$$

where $n$ and $c$ are the refractive index and the speed of light, respectively. The third-order nonlinear polarization, which is the lowest-order nonlinearity for isotropic materials, is given by

$$\vec{P}^{(3)}(t) = \tilde{\chi}^{(3)}(t)\vec{E}\vec{E}\vec{E},\tag{2}$$

where the proportionality factor $\tilde{\chi}^{(3)}$ is the third-order susceptibility. Under stationary and quasi-stationary conditions, the lth component of $\vec{P}^{(3)}(t)$ for a parametric light-light interaction can be described by the electric fields at time t:

$$P_l^{(3)}(t) = \sum_{m,n,o=1}^{3} \chi_{lmno}^{(3)} E_m(t)E_n(t)E_o(t).\tag{3}$$

The electric field and the nonlinear polarization are conveniently written by their frequency components as

$$E_l(t) = \frac{1}{2}\sum_k \{E_l(\omega_k)e^{-i\omega_k t} + c.c.\}\tag{4}$$

$$P_l^{(3)}(t) = \frac{1}{2}\sum_k \left\{P_l^{(3)}(\omega_k)e^{-i\omega_k t} + c.c.\right\}.\tag{5}$$

With Eq. (3), $P_l^{(3)}(t)$ can be rewritten as

$$P_l^{(3)}(t) = \frac{1}{8} \sum_{m,n,o,\alpha,\beta,\gamma} \chi_{lmno}^{(3)}(-\omega_\delta; \omega_\alpha, \omega_\beta, \omega_\gamma) E_m(\omega_\alpha) E_n(\omega_\beta) E_o(\omega_\gamma), \qquad (6)$$

where the relation $\omega_\delta = \omega_\alpha + \omega_\beta + \omega_\gamma$ is used. Then, the frequency component $P_l^{(3)}(\omega_\delta)$ is found to be

$$P_l^{(3)}(\omega_\delta) = \sum_{m,n,o} \frac{d}{4} \chi_{lmno}^{(3)}(-\omega_\delta; \omega_\alpha, \omega_\beta, \omega_\gamma) E_m(\omega_\alpha) E_n(\omega_\beta) E_o(\omega_\gamma), \qquad (7)$$

where $d$ is the number of terms $E_m(\omega_\alpha) E_n(\omega_\beta) E_o(\omega_\gamma)$ that contribute to $P_l^{(3)}(\omega_\delta)$. It is 6 when all three frequencies $\omega_\alpha$, $\omega_\beta$, and $\omega_\gamma$ are different. $d = 3$ if two of them are equal. $D = 1$ if two of them are equal, and d = 1 is the case for three identical frequencies.

Hereafter, a two-beam SRS process where the pump and Stokes beams represent the high- and low-frequency fields will be considered. Their frequencies are denoted as $\omega_p$ and $\omega_S$, respectively. When their beat (difference) frequency, $\omega_p - \omega_S$, becomes identical to the vibrational frequency $\omega_0$ of Raman-active mode in the medium, the stimulated Raman gain (loss) of the Stokes (pump) beam occurs. The corresponding Raman susceptibility, denoted as $\chi_R^{(3)}$, is well-known, i.e.

$$\chi_R^{(3)}(-\omega_S; \omega_p, -\omega_p, \omega_S) = \frac{1}{4m}\left(\frac{\partial\alpha}{\partial q}\right)^2 \frac{N(1 - 2n_{ex})}{\omega_0^2 - (\omega_p - \omega_S)^2 + 2i(\omega_p - \omega_S)^2/T_2} + \chi_{NR}^{(3)}, \quad (8)$$

where $m$, $\alpha$, $q$, $N$, $n_{ex}$, $T_2$, and $\chi_{NR}^{(3)}$ represent the reduced mass of vibrational mode, molecular polarizability, vibrational coordinate, the number density of Raman-active modes, the number of vibrational excited molecules, dephasing time, and non-resonant susceptibility. The latter, $\chi_{NR}^{(3)}$, is usually real and positive [28, 29]. Note that the imaginary part of Raman susceptibility $\chi_R^{(3)}$ is negative, which is important in understanding the Raman gain (loss) mechanism of the Stokes (pump) beam [24].

The present work considers the SRS process within paraxial approximation. The electric fields propagating along the $z$ direction can thus be written as, for $j$ = pump and Stokes,

$$E_j(x, y, z) = \mathcal{E}_j(x, y, z) e^{ik_j z - i\omega_j t}. \qquad (9)$$

The remaining problem is to obtain the spatial amplitudes $\mathcal{E}_j(x, y, z)$ and intensity distributions $\frac{1}{2}\varepsilon_0 cn|\mathcal{E}_j(x, y, z)|^2$ of the pump and Stokes beams when their input beam profiles are specifically designed using spiral phase plates, holographic gratings, metamaterials, or spatial light modulators [30]. Here, $\varepsilon_0$ represents the vacuum permittivity.

Inserting Eq. (9) into the wave equation (1) and invoking the slowly varying envelop approximation (SVEA),[26] i.e., $\frac{\partial^2 \mathcal{E}_j}{\partial z^2} \ll 2ik_j\left(\frac{\partial \mathcal{E}_j}{\partial z}\right)$ where $k_j$ is the wavenumber, one can find

$$\nabla_t^2 \mathcal{E}_p + 2ik_p\left(\frac{\partial \mathcal{E}_p}{\partial z}\right) = -\frac{4\pi k_p^2}{n_p^2} \chi_R^{(3)*} |\mathcal{E}_S|^2 \mathcal{E}_p \qquad (10)$$

$$\nabla_t^2 \mathcal{E}_S + 2ik_S \left( \frac{\partial \mathcal{E}_S}{\partial z} \right) = -\frac{4\pi k_S^2}{n_S^2} \chi_R^{(3)} |\mathcal{E}_p|^2 \mathcal{E}_S, \tag{11}$$

where

$$\nabla_t^2 \equiv \frac{\partial^2}{\partial x^2} + \frac{\partial^2}{\partial y^2}. \tag{12}$$

Note that the coefficient on the right-hand side of the wave equation for the pump (Stokes) beam contains $\chi_R^{(3)^*}$ ($\chi_R^{(3)}$). As the imaginary part of $\chi_R^{(3)}$ is negative (Eq. (8)), the pump (Stokes) beam intensity is reduced (increased) by the SRS process. It is evident that the two beams are coupled through the SRS process. The amplitudes, phases, and, consequently, intensity distributions of these beams are mutually dependent, suggesting the possibility of structuring or shaping the beam profile of either the pump or Stokes beam by controlling the other.

### 2.2. Laguerre-Gauss mode expansions of the pump and Stokes beams

The cases where the intensity distributions of the pump and Stokes beams exhibit cylindrical symmetry will be considered in this section. Consequently, the amplitudes of the pump and Stokes beams can generally be expanded in a complete set of orthonormal LG modes in a cylindrical coordinate system, represented as follows: [31, 32]

$$\mathcal{E}_p(r, \phi, z) = \sum_{n,l}^{\square} P_n^l(z) \, U_n^l(r, \phi, z) \tag{13}$$

$$\mathcal{E}_S(r, \phi, z) = \sum_{n,l}^{\square} S_n^l(z) \, V_n^l(r, \phi, z). \tag{14}$$

Here, $P_n^l(z)$ and $S_n^l(z)$ denote the complex mode amplitudes of the pump and Stokes beams, respectively. In Eqs. (13) and (14), $U_n^l(r, \phi, z)$ and $V_n^l(r, \phi, z)$ represent the LG modes, with $n$ and $l$ denoting the radial and rotational (azimuthal) mode indices, respectively. In the fields of telecommunication, optical cryptography, and imaging systems, $l$ is related to the topological charge of an optical vortex. Throughout this paper, it will be referred to as the rotational index.

The LG modes, $U_n^l(r, \phi, z)$ and $V_n^l(r, \phi, z)$, are the specific solutions of free-space wave equations for the pump and Stokes beams that are, for $j$ = pump and Stokes,

$$\nabla_t^2 \mathcal{E}_j + 2ik_j \left( \frac{\partial \mathcal{E}_j}{\partial z} \right) = 0. \tag{15}$$

The LG modes for the pump are [33]

$$U_n^l(r, \phi, z) = \frac{1}{w_p} \left[ \frac{2n!}{\pi(n + |l|)!} \right]^{1/2} \left[ \frac{\sqrt{2}r}{w_p} \right]^{|l|} L_n^l \left( \frac{2r^2}{w_p^2} \right) exp \left( -\frac{r^2}{w_p^2} - ik_p \frac{r^2}{2R_p} - il\phi + i\psi_p(z) \right) \tag{16}$$

Here, the beam waists of the pump and Stokes beams are determined by the corresponding confocal parameter and wavenumber as (for $m = p$ and $S$),

$$w_m^2 = \frac{2z_R^m}{k_m}\left(1 + \left(\frac{z}{z_R^m}\right)^2\right),$$  (17)

where $z_R^m$ (for $m = p$ and $S$) is the Rayleigh range or confocal parameter. $L_n^l(x)$ are the associated Laguerre polynomials. The Gouy phase is defined as

$$\psi_p(z) = (2n + |l| + 1)\tan^{-1}\left(\frac{z}{z_R^p}\right).$$  (18)

The phase front radius $R_p(z)$ of the pump is defined as

$$R_p(z) = z_R^p\{(z/z_R^p) + (z_R^p/z)\}.$$  (19)

The LG modes $V_n^l(r, \phi, z)$ for the Stokes beam is obtained by replacing $w_p$, $k_p$, and $z_R^p$ with $w_S$, $k_S$, and $z_R^S$, respectively, in Eq. (16).

Since the unit of LG modes is in 1/m and that of electric field amplitude is V/m, the units of $P_n^l(z)$ and $S_n^l(z)$ that are the complex mode amplitudes of the pump and Stokes beams are in V.

First, the LG-mode-expanded $\mathcal{E}_p$ and $\mathcal{E}_S$ in Eqs. (13) and (14) are inserted into the coupled wave equations (Eqs. (10) and (11)). Second, multiplying $U_{n'}^{l'^*}\left(V_{n'}^{l'^*}\right)$ and integrating over the transverse coordinates, $r$ and $\phi$, one can find the following coupled equations for the complex mode expansion coefficients:

$$\frac{d}{dz}P_{n'}^{l'}(z) = \sum_{n,l}\Lambda_{n',n}^{l',l}(z)P_n^l(z)$$  (20)

$$\frac{d}{dz}S_{n'}^{l'}(z) = \sum_{n,l}\Gamma_{n',n}^{l',l}(z)S_n^l(z),$$  (21)

where the $(l', l, n', n)$-dependent super-matrix elements inducing mode mixings are defined as

$$\Lambda_{n',n}^{l',l}(z) = i\frac{2\pi k_p}{n_p^2}\chi_R^{(3)^*}\int_0^{2\pi}d\phi\int_0^\infty dr\, r\, U_{n'}^{l'^*}\left|\sum_{p,q}S_p^q V_p^q\right|^2 U_n^l$$  (22)

$$\Gamma_{n',n}^{l',l}(z) = i\frac{2\pi k_S}{n_S^2}\chi_R^{(3)}\int_0^{2\pi}d\phi\int_0^\infty dr\, r\, V_{n'}^{l'^*}\left|\sum_{p,q}P_p^q U_p^q\right|^2 V_n^l.$$  (23)

The coefficient on the right-hand-side of Eq. (23), $\frac{2\pi k_S}{n_S^2}\chi_R^{(3)}$, is related to the Raman gain coefficient $g$ ($> 0$) of the plane wave as

$$g = -2Im\left[\frac{2\pi k_S}{n_S^2}\chi_R^{(3)}\right].$$ (24)

The Raman loss of the pump is determined by $-g\frac{k_p n_S^2}{k_S n_p^2}$. The unit of Raman susceptibility is m²/V². Therefore, the unit of $g$ is m/V². Note that the unit of $\Lambda_{n',n}^{l',l}$ and $\Gamma_{n',n}^{l',l}(z)$ is 1/m.

In Eqs. (22) and (23), the real and imaginary components of $\Lambda_{n',n}^{l',l}$ ($\Gamma_{n',n}^{l',l}(z)$) introduce couplings in magnitudes and phases between the $LG_{nl}$ and $LG_{n'l'}$ modes of the pump (Stokes) beam. For the pump (Stokes) beam, $\Lambda_{n',n}^{l',l}(z)$ ($\Gamma_{n',n}^{l',l}(z)$) is the complex loss (gain) factor. In particular, the real part of each diagonal matrix element of $\bar{\Lambda}(z)$ ($\bar{\Gamma}(z)$) is a crucial determinant of the loss (gain) associated with the corresponding pump (Stokes) LG mode. However, owing to the nonzero off-diagonal matrix elements of $\Lambda(z)$ ($\Gamma(z)$), the initial pump (Stokes) beam evolves into a superposition of various higher-order LG modes with decreasing (increasing) amplitudes as it traverses a Raman medium.

The coupled equations in Eqs. (20) and (21) are nonlinear differential equations since the super-matrix elements of $\bar{\Lambda}(z)$ ($\bar{\Gamma}(z)$) for the propagation of $P_p^q(z)$ ($S_p^q(z)$) depend on $S_p^q(z)$ ($P_p^q(z)$). Consequently, solving them analytically is not feasible.

An advantage of expressing the two beams in terms of LG modes is evident. By employing a finite difference method with Eqs. (10) and (11), one can numerically compute the propagated pump and Stokes fields. The grid for such numerical calculations should be established in the three-dimensional Cartesian or cylindrical coordinate system. However, despite their nonlinearity, the same coupled differential equations in the 3D coordinate system are reduced to a set of coupled equations in one dimension along the z-axis.

Upon solving the coupled differential equations in Eqs. (20) and (21) with the z-dependent super-matrix $\bar{\Lambda}(z)$ ($\bar{\Gamma}(z)$) provided in Eq. (22) (Eq. (23)), the resulting $P_{n'}^{l'}(z)$ and $S_{n'}^{l'}(z)$ matrices, dependent on the radial and rotational indices $n'$ and $l'$, are employed to derive the amplitudes and phases of the pump and Stokes beams using Eqs. (13) and (14).

### 2.3. Cylindrically symmetric pump and Stokes beams

The coupled wave equations for the pump and Stokes beams, $\mathcal{E}_p(r,\phi,z)$ and $\mathcal{E}_S(r,\phi,z)$, were transformed into another set of coupled equations (Eqs. (20) and (21)) for $P_n^l(z)$ and $S_n^l(z)$, which depend only on $z$. This transformation was feasible because the two beams considered were expanded in terms of a complete set of LG modes that satisfy the free space wave equation in Eq. (15). In the present study, the loss and gain functions, which are proportional to $|\mathcal{E}_S(r,\phi,z)|^2$ and $|\mathcal{E}_p(r,\phi,z)|^2$, are assumed to be cylindrically symmetric. This assumption implies that the pump and Stokes beams are one of the LG modes or their linear combinations. Therefore, in the integrands of Eqs. (22) and (23), $\left|\sum_{p,q}S_p^q V_p^q\right|^2$ and $\left|\sum_{p,q}P_p^q U_p^q\right|^2$ are independent of $\phi$. As a result, the integration of the products of LG modes with different rotational indices, such as $U_{n'}^{l'*}U_n^l$ and $V_{n'}^{l'*}V_n^l$ in Eqs. (22) and (23), respectively, over $\phi$ vanishes if $l' \neq l$. Consequently, the coupled equations in Eqs. (20) and (21) become simplified as

$$\frac{d}{dz}P_{n'}^l(z) = \sum_{n,l}\Lambda_{n',n}^l(z)P_n^l(z)$$ (25)

$$\frac{d}{dz} S_{n'}^{l}(z) = \sum_{n,l} \Gamma_{n',n}^{l}(z)\, S_n^{l}(z). \tag{26}$$

If the initial intensity distributions of the pump and Stokes beams are cylindrically symmetric, the LG modes with different rotational indices $l'$ and $l$ remain uncoupled, even in cases where the SRS process occurs. For example, consider the scenario where the initial pump beam is the TEM$_{00}$ mode, a paraxial Gaussian beam. In this case, only those LG modes with $l = 0$ become coupled and contribute to the resulting superposition state of the output pump beam. Similarly, if the initial Stokes beam is a doughnut beam, e.g., LG$_{01}$ mode, the output Stokes beam becomes a superposition of LG$_{n1}$ modes. The relative weighting factor $S_n^1(z)$ of the $n$th LG$_{n1}$ mode $V_n^l(r, \phi, z)$ for the propagated Stokes beam is primarily determined by (i) the Raman gain coefficient $g$, which is proportional to $-Im[\chi_R^{(3)}]$, (ii) pump beam intensity, and (iii) the mode mixing matrix elements $\Gamma_{n',n}^{1}(z)$.

While the coupled wave equations for the pump and Stokes beams are simplified into Eqs. (25) and (26), solving the coupled equations for the expansion coefficients remains challenging. Therefore, in the following sections, I will consider two approximate cases: (i) the stimulated Raman loss of the pump beam with an intense doughnut-shaped Stokes beam and (ii) the stimulated Raman gain of the Stokes beam with an intense depletion-free Gaussian pump beam, because they are of practical importance.

## 3. Stimulated Raman with Gaussian and doughnut beams: Decoupled equations

### 3.1 Theoretical description of SRL: Weak Gaussian pump and strong doughnut Stokes

The coupled equations for the pump and Stokes beams in Eqs. (10) and (11), or the corresponding equations for the expansion coefficients in Eqs. (25) and (26), can be decoupled when one of the two beams is much stronger than the other. In this subsection, I consider the scheme proposed in Ref. [19], where the intensity of the doughnut-shaped Stokes beam is assumed to be substantially stronger than that of the Gaussian pump, implying a situation of negligible gain of the Stokes beam. The expansion coefficients $S_n^l(z)$ for the Stokes beam remain constant throughout the Raman medium. If the incident Stokes beam is one of the LG modes, e.g., LG$_{01}$, then $\Gamma_{0,0}^1(z) \neq 0$ and all the other elements are $\Gamma_{n',n}^1(z) = 0$. Thus, it is necessary to determine the evolutions of the expansion coefficients $P_n^l(z)$ for the pump beam only.

To solve the differential equation for $P_n^l(z)$ given in Eq. (25), specifying the initial conditions is necessary. The LG$_{nl}$ modes with $l \geq 1$ exhibit doughnut-shaped intensity distributions with $n$ radial nodes. For simplicity, I will focus on cases where the incident Stokes beam is one of the LG$_{0l}$ modes. These modes have zero intensity on the optic axis and no radial node, i.e.,

$$\mathcal{E}_S(r, \phi, z_0) = S_0^{l_S}(z_0) V_0^{l_S}(r, \phi, z_0)$$

$$S_0^{l_S}(z_0) \neq 0$$

$$S_n^{l'}(z_0) = 0 \text{ for } n \neq 0 \text{ and } l' \neq l_S, \tag{27}$$

where $l_S$ is the rotational index of the initial Stokes LG mode, and $z_0$ is the $z$-position of the front surface of the Raman-active sample. The amplitude of the Stokes beam is determined by the initial value $S_0^{l_S}(z_0)$ in V. In this subsection, as the Stokes beam is assumed to be substantially stronger than the pump, resulting in negligible gained intensity after passing through the sample, the Stokes beam remains unchanged beyond the Raman medium, and $\frac{d}{dz} S_n^l(z) = 0$ for all $n$ and $l$.

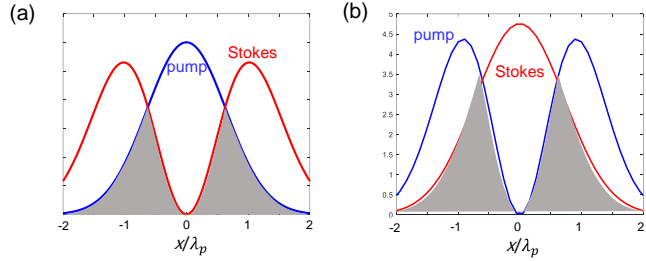

**Figure 2.** The intensity distributions of the key pump and Stokes beams used in the Cho (a) and Frontiera (b) schemes. Among three incident beams, pump, Stokes-1, and Stokes-2 (Figure 1(b)), the SRS-1 process by the Gaussian pump and doughnut-shaped Stokes-1 is the most critical step to achieve a super-resolution SRS or CARS microscopy. The shaded area in grey is the spatial region where the SRL of the pump occurs, whereas, in the central region where the Stokes beam intensity is low, the pump beam intensity remains unchanged. (b) In the Frontiera scheme, it is necessary to examine the beam profile of the Gaussian Stokes beam (Figure 1(a)) when its intensity is increased through the SRG process induced by the doughnut-shaped pump beam.

The incident pump beam at $z_0$ is assumed to be the $LG_{00}$ mode, i.e.,

$$\mathcal{E}_p(r, \phi, z_0) = P_0^0(z_0) U_0^0(r, \phi, z_0)$$

$$P_0^0(z_0) \neq 0$$

$$P_n^l(z_0) = 0 \text{ for } n \neq 0 \text{ and } l \neq 0. \tag{28}$$

The absolute amplitude of the pump beam is determined by the initial value of $P_0^0(z_0)$ in V. In Figure 2(a), the combination of a Gaussian pump beam and an $LG_{01}$ Stokes beam is illustrated. In the spatial region where the two beams overlap, the stimulated Raman loss process occurs, leading to changes in the intensity distribution of the pump. Theoretically, this is equivalent to the SRL-induced changes in the expansion coefficients $P_n^0(z_0)$, which are fully determined by Eq. (25). Hereafter, the superscript $l$ in Eq. (25), which is zero for the initial condition of the pump beam given in Eq. (28), will be omitted. Thus, the differential equation in Eq. (25) can be recast in linear algebraic form as

$$\frac{d}{dz} \vec{P}(z) = \bar{\Lambda}(z) \vec{P}(z), \tag{29}$$

where $\vec{P}(z)$ is a column vector with elements $P_n^0(z)$ and $\bar{\Lambda}(z)$ is an $N$ by $N$ matrix with $N \, LG_{n0}$ basis modes ($0 \leq n \leq N-1$).

The matrix elements of $\bar{\Lambda}(z)$ is given as:

$$\Lambda_{n',n}^{\square}(z) = i\frac{2\pi k_p}{n_p^2}\chi_R^{(3)^*}|S_0^{ls}(z_0)|^2\int_0^{2\pi}d\phi\int_0^{\infty}dr\, r\,\left|V_0^{ls}(r)\right|^2 U_{n'}^{0\,^*}(r,z)U_n^0(r,z), \quad (30)$$

where the two terms in the integrand in the above equation are

$$\left|V_0^{ls}(r)\right|^2 = \frac{2}{\pi}\frac{1}{|l_S|!}\frac{1}{w_S^2}\left(\frac{2r^2}{w_S^2}\right)^{|l_S|}exp\left(-\frac{2r^2}{w_S^2}\right)\left(L_0^{|l_S|}\left(\frac{2r^2}{w_S^2}\right)\right)^2 \quad (31)$$

$$U_{n'}^{0\,^*}(r,z)U_n^0(r,z) = \frac{2}{\pi}\frac{1}{w_p^2}\,exp\left(-\frac{2r^2}{w_p^2}\right)L_{n'}^0\left(\frac{2r^2}{w_p^2}\right)L_n^0\left(\frac{2r^2}{w_p^2}\right)e^{-2i(n'-n)\tan^{-1}\left(\frac{z}{z_R^p}\right)}. \quad (32)$$

Since the matrix $\bar{\Lambda}(z)$ depends on $z$, Eq. (29) is not an ordinary linear differential equation. However, it can be further simplified as follows. Firstly, I assume that the confocal parameters of the pump and Stokes beams are approximately the same or can be made to be the same by controlling the microscope objective lenses, i.e.,

$$z_R = z_R^p = z_R^S. \quad (33)$$

Usually, the frequencies of the pump and Stokes beams differ from each other by the amount of vibrational photon energy; this is a good and acceptable quantitative approximation. Later in Sec. IV, I will solve the pump-Stokes coupled equations in Eqs. (25) and (26) numerically, employing a finite-difference method, where the above approximation in Eq. (33) will not be invoked. However, for the present approximate calculations in the limit that the Stokes gain is negligible, the approximation in Eq. (33) is useful. Secondly, a new dimensionless variable $\rho$ associated with the radial coordinate $r$ is defined as

$$\rho = 2r^2\left[\frac{1}{w_p^2} + \frac{1}{w_S^2}\right]. \quad (34)$$

Introducing dimensionless $\sigma_m$ for $m = p$ and $S$, which is a measure of the spatial overlap of the pump and Stokes beam intensity distributions, as

$$\sigma_m \equiv \frac{k_m}{k_p + k_S}, \quad (35)$$

one can find that

$$2r^2/w_p^2 = \sigma_p\rho$$

$$2r^2/w_S^2 = \sigma_S\rho = (1-\sigma_p)\rho$$

$$r\,dr = \frac{\sigma_p w_p^2}{4}d\rho. \quad (36)$$

Using this newly introduced variable $\rho$ and the pump beam overlap parameter $\sigma_p$, the matrix elements $\Lambda_{n',n}^{\square}(z)$ can be rewritten as

$$\Lambda_{n',n}^{\square}(z) = i \frac{2 \, k_p k_S \chi_R^{(3)^*} \left| S_0^{l_S}(z_0) \right|^2 \sigma_p}{n_p^2 \, |l_S|!} \frac{\exp\left\{-2i(n'-n)\tan^{-1}\left(\frac{z}{z_R}\right)\right\}}{z_R \left(1 + \left(\frac{z}{z_R}\right)^2\right)} f_{n',n}^{\square}(\sigma_p), \quad (37)$$

where $\sigma_p$-dependent $f_{n',n}^{\square}$, which is a real number and independent of $z$, is defined as

$$f_{n',n}^{\square}(\sigma_p) = \int_0^\infty d\rho \, e^{-\rho} \left[(1-\sigma_p)\rho\right]^{|l_S|} \left(L_p^{|l_S|}\left[(1-\sigma_p)\rho\right]\right)^2 L_{n'}^0(\sigma_p\rho) L_n^0(\sigma_p\rho). \quad (38)$$

The integral in Eq. (38) is to be calculated numerically. In the middle of Eq. (37), there exist terms that are dependent on $z$. Therefore, thirdly, another coordinate $\zeta$ associated with $z$ is defined as

$$\zeta = \tan^{-1}\left(\frac{z}{z_R}\right). \quad (39)$$

Then, $dz = z_R \left(1 + \left(\frac{z}{z_R}\right)^2\right) d\zeta$. By substituting Eq. (39) into (37) and rewriting the differential equation of the $z$ coordinate into that of the $\zeta$ coordinate, Eq. (29) can be recast in the following form:

$$\frac{d}{d\zeta}\vec{P}(\zeta) = e^{-i\bar{h}\zeta}\, \bar{L}\, e^{i\bar{h}\zeta}\vec{P}(\zeta), \quad (40)$$

where $\bar{h}$ is the diagonal matrix defined as

$$\bar{h} = \begin{pmatrix} 0 & 0 & 0 & \cdots \\ 0 & 2 & 0 & \cdots \\ 0 & 0 & 4 & \cdots \\ \vdots & \vdots & \vdots & \ddots \end{pmatrix}. \quad (41)$$

The diagonally symmetric matrix $\bar{L}$ has the following dimensionless complex-valued elements:

$$L_{n',n}^{\square} = i \frac{2k_p k_S \chi_R^{(3)^*} \left| S_0^{l_S}(z_0) \right|^2 \sigma_p}{n_p^2 |l_S|!} f_{n',n}^{\square}(\sigma_p). \quad (42)$$

Finally, the solution of Eq. (40) is obtained:

$$\vec{P}(\zeta) = e^{-i\bar{h}\zeta}\, exp\left\{(\bar{L} + i\bar{h})(\zeta - \zeta_0)\right\} e^{i\bar{h}\zeta_0}\, \vec{P}(\zeta_0), \quad (43)$$

where $\zeta_0 = \tan^{-1}\left(\frac{z_0}{z_R}\right)$.

Since the imaginary part of the Raman susceptibility $\chi_R^{(3)}$ is a negative real number and the diagonal element $f_{0,0}^{\square}$ is a positive real number, the diagonal matrix element $L_{0,0}^{\square}$ is a real negative number. Thus, aside from the phase modulation resulting from the nonzero real part of the Raman susceptibility $\chi_R^{(3)}$, the amplitude of the pump beam decreases with an increase in the interaction length, $\zeta - \zeta_0$, through a given Raman medium. This reflects the well-known stimulated Raman loss of the pump beam via the SRS process. However, what is new and important is that the contributions from other higher-order $LG_{n0}$ modes to the propagated pump beam become significant. The propagator matrix, $e^{-i\hbar\zeta} exp\{(\bar{L} + i\bar{h})(\zeta - \zeta_0)\} e^{i\hbar\zeta_0}$, has nonzero off-diagonal elements. Therefore, even with the pump beam with initial conditions in Eq. (29), the expansion coefficients $P_n^0(z)$ $(z > z_0)$ for higher-order $LG_{n0}$ modes become nonzero. Note that the beam waist of the central lobe of higher-order $LG_{n0}$ mode is narrower than that of the fundamental Gaussian mode. Consequently, the pump beam is expected to undergo focusing via the SRL process induced by an intense doughnut-shaped Stokes beam. This expected focusing phenomenon differs from Kogelnik's gain focusing for a uniform parabolic distribution [34].

### 3.2 Numerical simulation of SRL: Weak Gaussian pump and strong doughnut Stokes

Since the formal solution of the differential equation in Eq. (43) was obtained, it becomes possible to calculate the propagation of LG mode expansion coefficients, $P_n^0(z)$. They are used to predict the $z$-dependent amplitude and phase of the pump beam, $\mathcal{E}_p(r, \phi, z) = \sum_{n,l}^{\square} P_n^0(z) U_n^0(r, \phi, z)$. The numerical calculations involved in this process necessitate the incorporation of experimental setup details.

**Table 1.** Parameters used in the numerical calculations of the SRL pump and SRG Stokes beams.

| parameters | |
|---|---|
| $\lambda_p$ | 800 nm |
| $\lambda_S$ | 1000 nm |
| $\sigma_p$ | 0.556 |
| $\sigma_S$ | 0.444 |
| NA | 0.3 |
| $n_p^{\square} = n_S^{\square} \ (water)$ | 1.33 |
| $z_R^p$ | 3.70 μm |
| $z_R^S$ | 4.62 μm |
| $z_R$ | 4.16 μm |
| $L$ | $10z_R = 41.6$ μm |
| $g$ | $2.489 \times 10^{-18}$ m/V$^2$ |
| $-Im[\chi_R^{(3)}]$ | $5.58 \times 10^{-26}$ m$^2$/V$^2$. |

Table 1 provides a summary of the wavelengths of the two beams and other microscope parameters. In Figure 3, the fundamental Gaussian beam profile of the pump, including the confocal parameter, beam waist ($w_p$), and the thickness of the Raman medium, is illustrated. The origin of the laboratory coordinate frame is assumed to be at the focus of the fundamental Gaussian pump beam propagating from left to right. The front surface of the Raman-active sample is positioned away from the origin by $L$ at $z_0$ $(= -L)$. The average confocal parameter (Rayleigh range) $z_R$ is set to be 4.16 μm. By utilizing the wavelengths of the pump and Stokes beams, the beam overlap parameter $\sigma_p$, as defined in Eq. (35), is determined to be 0.556.

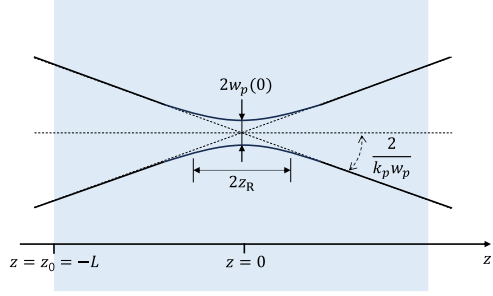

**Figure 3.** Focused pump beam profile under paraxial approximation. The radius of the beam at the focus is $w_p(z = 0)$.[33] The Rayleigh range, often called the confocal parameter, is denoted as $z_R$. The $z$-position of the front surface of the Raman medium is at $-L$.

For the sake of completeness, it is essential to note that the dimensions of $\mathcal{E}_m^0$, $P_n^\square$ (and $S_n^\square$), $U_p^l$ (and $V_p^l$), $\chi_R^{(3)}$, $g$, and $\Lambda_{n',n}^\square$ (and $\Gamma_{n',n}^\square$) are given in units of V/m, V, 1/m, m²/V², m/W (or m/V²), and 1/m respectively. The typical beam intensity $I$ ($= \frac{1}{2}\varepsilon_0 c n |\mathcal{E}|^2$) for coherent Raman imaging is a few mW/μm² (= nW/nm²), and the Raman gain coefficient $g$ of a plane wave ranges from $10^{-15}$ to $10^{-11}$ m/W.

To convert the beam intensity into the field amplitude, one can employ $\varepsilon_0 = 8.854 \times 10^{-12}$ F/m and $c = 2.998 \times 10^8$ m/s. Therefore, $I$ in W/m² is related to $|\mathcal{E}|$ as $1.327 \times 10^{-3} n |\mathcal{E}|^2$, where $|\mathcal{E}|$ is in V/m. A beam intensity of 1 mW/μm² corresponds to approximately $0.752$ MV/m for the electric field amplitude $|\mathcal{E}|$ in water with $n_w = 1.33$.

Hereafter, it is assumed that a typical Raman gain coefficient at the Stokes beam wavelength (1 μm) to be $0.9 \times 10^{-15}$ m/W. In coherent Raman imaging studies of cells, such as those focusing on proteins and lipids, the concentrations of typical biomolecular materials are typically low, resulting in Raman gain coefficients smaller than this assumed value. In the case of a Raman medium comprising water, the Raman gain coefficient for OH stretch vibrations falls within the range of $10^{-13}$ to $10^{-14}$ m/W [35, 36]. Notably, coefficients for other organic solvents exceed that of water. In the subsequent subsections, numerical calculation results for typical materials with dimensionless gain and loss coefficients are presented. It should be mentioned that the absolute magnitude of the gain coefficient is not critical for the present numerical calculation studies.

For completeness, it is possible to convert the assumed value of $g$ ($= 0.9 \times 10^{-15}$ m/W) into $2.489 \times 10^{-18}$ m/V² using the factor $\frac{1}{2}\varepsilon_0 c n_w$. Utilizing the definition of $g$ in Eq. (24), the refractive index of water, and the wavelength of Stokes beam, one can estimate $-Im[\chi_R^{(3)}]$ for this material to be $5.58 \times 10^{-26}$ m²/V². It is widely recognized that the Raman gain coefficient varies widely depending on materials, temperature, and the wavelengths of the pump and Stokes beams, rendering the above estimates as illustrative examples.

Consider a scenario where the pump takes the form of a fundamental Gaussian beam, and the Stokes adopts a doughnut-shaped $LG_{0l}$ mode with $l \geq 1$ (Figures 1(b) and 2(a)). The spatial region (highlighted in grey in Figure 2(a)), where the intensity distributions of the pump and Stokes beam overlap, becomes the site for the SRL process in the pump beam. In contrast, the Stokes beam intensity experiences an increase in intensity within the same region due to the SLG process.

In section III, the assumption that the intensity of the Stokes beam significantly surpasses that of the pump was used, and as a result, the SRG of the Stokes beam during its propagation can be neglected. Nevertheless, the SRL process induced by the spatially non-homogeneous Stokes-doughnut beam induces changes in the amplitude and phase of the pump beam as it traverses through the Raman medium.

For numerical calculations, the following dimensionless quantity is introduced as the descriptive variable representing the extent of SRL of the pump:

$$i\frac{2k_p k_S {\chi_R^{(3)}}^* |S_0^{l_S}(z_0)|^2 \sigma_p}{n_p^2 |l_S|!} = loss \times (-1 + 0.1i).$$ (44)

Here, $loss$ ($\geq 0$) is regarded as the adjustable variable for subsequent numerical calculations. Considering the estimates and parameters outlined in Table 1, the $loss$ is approximately of the order of unity, specifically 0.978. Experimentally, the $loss$ for a given material can be increased by increasing the Stokes beam intensity, which is proportional to $|S_0^{l_S}(z_0)|^2$ in Eqs. (42) and (44), reaching values on the order of MV. The negative sign in "$-1$" on the right-hand side of

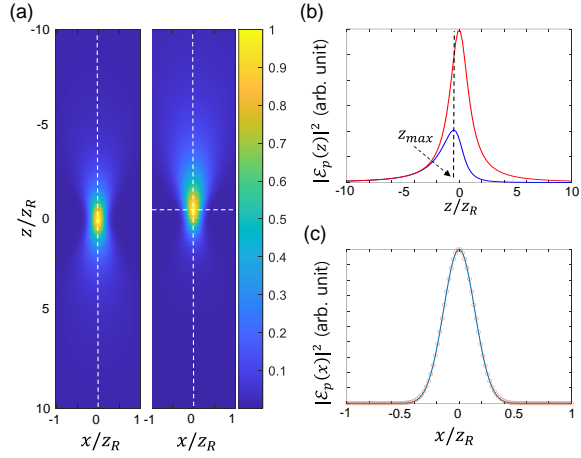

**Figure 4.** The intensity distributions of the Gaussian pump beam in the absence and presence of a doughnut-shaped Stokes beam with an intensity similar to that of the pump. (a) The intensity distributions of the pump beam on the $x$-$z$ plane without (left) and with (right) the doughnut-shaped Stokes beam. (b) The pump beam intensity at $x = 0$ and $y = 0$ is plotted with respect to $z/z_R$. The red and blue lines correspond to those without and with the doughnut-shaped Stokes beam, respectively. The peak maximum position of the SRL pump beam intensity is denoted as $z_{max}$. (c) The pump beam intensity at $z = z_{max}$ is plotted with respect to $x/z_R$. The red open circle and blue line correspond to the pump beam intensities without and with the doughnut-shaped Stokes beam, respectively.

Eq. (44) emphasizes that the pump experiences a loss in its intensity. Despite the susceptibility $\chi_R^{(3)}$ being purely imaginary at a vibrational resonance, a small imaginary part, $0.1i$, owing to the non-resonant contribution $\chi_{NR}^{(3)}$ was included. This addition, though not impacting the SRL, is included for the sake of completeness.

In the Cho scheme, it was proposed that the pump beam would undergo narrowing as it traverses a given Raman medium through the SRL of the pump when the conjugate Stokes beam has a doughnut-shaped profile. Initially, utilizing the typical parameters summarized in Table 1, including $|P_0^0(z_0)| = |S_0^{l_S=1}(z_0)| = 0.752$ MV, one can analyze the evolution of the weak Gaussian pump beam along the z-axis. The objective is to explore the possibility of the Gaussian-shaped pump beam gradually narrowing as its peripheral region is trimmed away. For comparison, the Gaussian pump beam intensity distribution in the absence of the Stokes beam is also shown in Figure 4(a) (left panel). The SRL pump intensity distribution on the $x$-$z$ plane is shown in Figure 4(a) (right panel). Due to the SRL of the pump, the focal point along the $z$-axis shifts toward the objective lens (upper end of the panel) at $z = -0.5z_R$. To depict the SRL process, the peak intensity of the pump beam with respect to $z$ is plotted (blue line, Figure 4(c)). In the vicinity of the pump beam focus at $z = 0$, the SRL is substantial (compared to the red circles in Figure 4(c)), and its intensity diminishes at $z = 4z_R$. At the peak position $z_{max} = -0.5z_R$, the normalized pump intensity distributions with and without the SRL process induced by the Stokes beam with respect to $x$ are plotted in Figure 4(c). The corresponding beam widths are quantitatively similar, suggesting that the doughnut Stokes-induced SRL of the pump does not significantly contribute to beam narrowing.

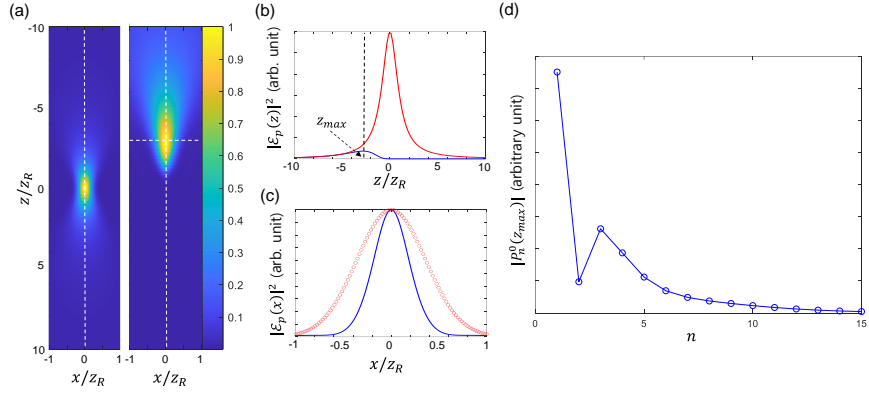

**Figure 5.** The intensity distributions of the Gaussian pump beam in the absence and presence of a doughnut-shaped Stokes beam with an intensity approximately ten times stronger than that of the pump. (a) The intensity distributions of the pump beam on the $x$-$z$ plane without (left) and with (right) the doughnut-shaped Stokes beam. (b) The pump beam intensity at $x = 0$ and $y = 0$ is plotted with respect to $z/z_R$. The red and blue lines correspond to those without and with the doughnut-shaped Stokes beam, respectively. (c) The pump beam intensity distribution at $z = z_{max}$ is plotted with respect to $x/z_R$. The red open circle is the pump beam intensity without the Stokes beam, whereas the blue line corresponds to the pump beam intensity in the presence of the doughnut-shaped Stokes beam. (d) The absolute magnitudes of $|P_n^0(z_{max})|$ are plotted for different $n$'s.

Next, let us examine the case that the incident Stokes beam intensity is increased by a factor of ten, specifically $|P_0^0(z_0)| = 0.752$ MV and $|S_0^{ls}(z_0)| = 0.752\sqrt{10}$ MV, leading to a corresponding increase in the *loss* value by the same factor. In Figure 5(a) (right panel), the pump beam intensity distribution on the $x$-$y$ plane is depicted. Due to the pronounced SRL effect (red line compared to the blue line in Figure 5(b)), the peak position of the pump along the $z$-axis is at $z_{max} = -2.7z_R$, with the intensity distribution along the $x$-axis shown in Figure 5(c). Notably, the width of the SRL pump beam (blue line in Figure 5(c)) is much smaller than that of the pump beam (red circles in Figure 5(c)) in the absence of the Stokes beam. This outcome demonstrates that the doughnut-shaped Stokes beam narrows the pump beam through stimulated Raman coupling.

In order to understand the underlying mechanism behind the SRL-induced narrowing of the Gaussian pump beam by a strong doughnut-shaped Stokes beam, the values of $|P_n^0(z)|$ at the peak of $z_{max} = -2.7z_R$ are plotted in Figure 5(d). These values represent the expansion coefficients of the radial LG modes, $U_n^0(r, \phi, z)$. Although the initial conditions are $|P_{n=0}^0(z_0)| \neq 0$ and $P_{n>0}^0(z_0) = 0$, due to the SR-induced mode-mixing processes, the expansion coefficients $P_{n>0}^0(z)$ become nonzero, making the contributions from $U_{n>0}^0(r, \phi, z)$ to the pump beam profile significant. Considering that the central lobe of $U_{n>0}^0(r, \phi, z)$ is narrower than the fundamental $LG_{00}$ mode (Figure 6), $U_0^0$, the resulting pump beam becomes narrower than the pump beam without the SRL process.

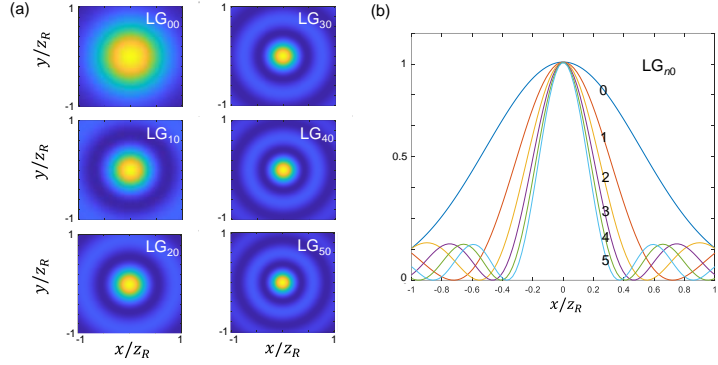

**Figure 6.** (a) The pump intensity distributions of $LG_{n0}$ mode on the *x-y* plane at the focus. (b) The pump intensity distributions are plotted with respect to $x/z_R$.

Nevertheless, a limitation of the Cho scheme is evident in that the focal point of the SRL pump significantly deviates from the beam focus of the third Stokes beam employed to acquire SRS or CARS microscopic images. As a consequence of the nearly complete loss of the pump beam intensity at $z = 0$, the resulting SRS or CARS microscopic signal generated may be exceedingly weak. This weakness poses a challenge to the practical application of the Cho scheme.

### 3.3 Theoretical description of SRG: Strong doughnut pump and weak Gaussian Stokes

In this subsection, the approach proposed by Silva et al. (Figure 1(a)) is explored. In this context, I assume that the intensity of the doughnut-shaped pump beam is substantially stronger than that of the Gaussian Stokes beam, representing a scenario of a pump beam with minimal depletion. Here, our focus is on measuring the SRG of the Gaussian Stokes beam. Given that the incident pump beam is one of the toroidal LG modes, such as $LG_{0l}$, where $\Lambda_{0,0}^l(z) \neq 0$ and all the other elements $\Lambda_{n',n}^l(z)$ are zero, it is necessary to determine the evolution of the expansion coefficients $S_n^l(z)$ for the Stokes beam.

The initial conditions for the incident pump beam are given as

$$\mathcal{E}_p(r, \phi, z_0) = P_0^{l_p}(z_0) U_0^{l_s}(r, \phi, z_0)$$

$$P_0^{l_p}(z_0) \neq 0$$

$$P_n^{l'}(z_0) = 0 \text{ for } n \neq 0 \text{ and } l' \neq l_s, \qquad (45)$$

where $l_p$ is the rotational index of the initial pump LG mode. The incident Stokes beam at $z_0$ is assumed to be the $LG_{00}$ mode, i.e.,

$$\mathcal{E}_S(r, \phi, z_0) = S_0^0(z_0) V_0^0(r, \phi, z_0)$$

$$S_0^0(z_0) \neq 0$$

$$S_n^l(z_0) = 0 \text{ for } n \neq 0 \text{ and } l \neq 0. \tag{46}$$

In the spatial region where the two beams overlap, the stimulated Raman gain of the Stokes beam occurs, while in the region close to the optic axis, the SRS is weak. The following differential equation for $\vec{S}(z)$, a column vector with elements $S_n^0(z)$, needs to be solved:

$$\frac{d}{dz}\vec{S}(z) = \bar{\Gamma}(z)\vec{S}(z), \tag{47}$$

where $\bar{\Gamma}(z)$ is an $N$ by $N$ matrix with $N$ LG$_{n0}$ basis modes ($0 \leq n \leq N-1$). The matrix elements of the gain matrix $\bar{\Gamma}(z)$:

$$\Gamma_{n',n}^{\square}(z) = i\frac{2\pi k_S}{n_S^2}\chi_R^{(3)}|P_0^{lp}(z_0)|^2 \int_0^{2\pi}d\phi \int_0^{\infty}dr\, r\, \left|U_0^{lp}(r)\right|^2 V_{n'}^{0\,*}(r,z)V_n^0(r,z), \tag{48}$$

where the two terms in the integrand in the above equation are

$$\left|U_0^{lp}(r)\right|^2 = \frac{2}{\pi}\frac{1}{|l_p|!}\frac{1}{w_p^2}\left(\frac{2r^2}{w_p^2}\right)^{|l_p|}exp\left(-\frac{2r^2}{w_p^2}\right)\left(L_0^{|l_p|}\left(\frac{2r^2}{w_p^2}\right)\right)^2 \tag{49}$$

$$V_{n'}^{0\,*}(r,z)V_n^0(r,z) = \frac{2}{\pi}\frac{1}{w_S^2}\,exp\left(-\frac{2r^2}{w_S^2}\right)L_{n'}^0\left(\frac{2r^2}{w_S^2}\right)L_n^0\left(\frac{2r^2}{w_S^2}\right)e^{-2i(n'-n)\tan^{-1}\left(\frac{z}{z_R^S}\right)}. \tag{50}$$

Again the approximation $z_R = z_R^p = z_R^S$ is invoked. The dimensionless variable $\rho$ in Eq. (34) and the spatial overlap of the pump and Stokes beam intensity distributions $\sigma_S$ are used to obtain

$$\Gamma_{n',n}^{\square}(z) = i\frac{2k_p k_S \chi_R^{(3)}\left|P_0^{lp}(z_0)\right|^2 \sigma_S}{n_S^2\,|l_p|!}\frac{\exp\left\{-2i(n'-n)\tan^{-1}\left(\frac{z}{z_R}\right)\right\}}{z_R\left(1+\left(\frac{z}{z_R}\right)^2\right)}f_{n',n}^{\square}(\sigma_S), \tag{51}$$

where $f_{n',n}^{\square}(\sigma)$ was defined in Eq. (38). Using the coordinate $\zeta$ in Eq. (39) and following the same line of derivation to obtain Eq. (43), one can find the solution for $\vec{S}(\zeta)$:

$$\vec{S}(\zeta) = e^{-i\bar{h}\zeta}\,exp\left\{(\bar{G}+i\bar{h})(\zeta-\zeta_0)\right\}e^{i\bar{h}\zeta_0}\,\vec{S}(\zeta_0), \tag{52}$$

where the elements of the SRG matrix $\bar{G}$ are

$$G_{n',n}^{\square} = i\frac{2k_p k_S \chi_R^{(3)}\left|P_0^{lp}(z_0)\right|^2 \sigma_S}{n_S^2|l_p|!}f_{n',n}^{\square}(\sigma_S). \tag{53}$$

Noting that the imaginary part of the Raman susceptibility $\chi_R^{(3)}$ is a negative real number and the diagonal element $f_{0,0}^{\square}$ is a positive real number, one can confirm the positivity of the diagonal matrix element $G_{0,0}^{\square}$, indicating the occurrence of the SRG process. As the Gaussian Stokes beam propagates through the Raman medium, the contribution of higher-order LG$_{n0}$ modes with $n > 0$ becomes significant in the transmitted Stokes beam. The critical question is whether this interplay between a strong doughnut-shaped pump and a weak Gaussian Stokes results in the narrowing of the Stokes beam through SRG. Numerical calculation results addressing this will be presented in the following subsection D.

### 3.4 Numerical simulation of SRG: Strong doughnut pump and weak Gaussian Stokes

The parameters provided in Table 1 are used again. The average confocal parameter (Rayleigh range) $z_R$ is set to be 4.16 μm, and the beam overlap parameter $\sigma_S$ defined in Eq. (35) is 0.444. In this subsection, I assume that the doughnut-shaped ($l_p = 1$) pump beam intensity is much stronger than that of the Gaussian Stokes beam. This condition leads to the SRG process by the pump beam, causing changes in the amplitude and phase of the Stokes beam during its propagation through the Raman medium.

For numerical calculations, I consider the following dimensionless quantity as the descriptive variable determining the extent of stimulated Raman gain of the Stokes:

$$i \frac{2k_p k_S \chi_R^{(3)} |P_0^{l_p}(z_0)|^2 \sigma_S}{n_S^2 |l_p|!} = gain \times (1 + 0.1i). \tag{54}$$

Here, $gain$ ($\geq 0$) is an adjustable variable for the following numerical simulations. In Ref. [18] with the Frontiera scheme, they experimentally changed $gain$ for a given material, increasing the doughnut-shaped pump beam intensity that is proportional to $|P_0^{l_p}(z_0)|^2$. Again, the imaginary part on the right-hand side of Eq. (54) represents the non-resonant contribution $\chi_{NR}^{(3)}$ to $\chi_R^{(3)}$.

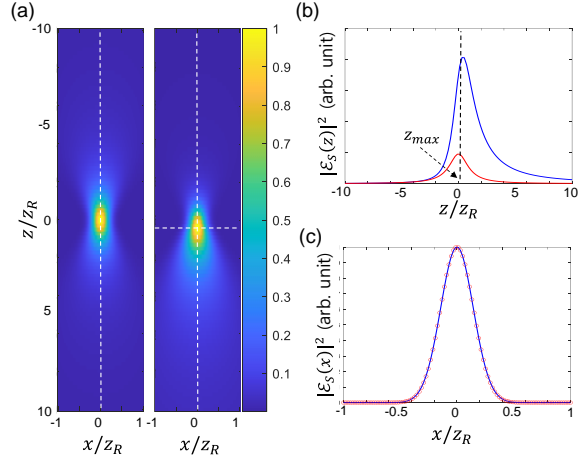

**Figure 7.** The intensity distributions of the Gaussian Stokes beam in the absence and presence of a doughnut-shaped pump beam with an intensity similar to that of the Stokes. (a) The intensity distributions of the Stokes beam on the $x$-$z$ plane without (left) and with (right) the doughnut-shaped pump beam. (b) The Stokes beam intensity at $x = 0$ and $y = 0$ is plotted with respect to $z/z_R$. The red and blue lines correspond to those without and with the doughnut-shaped Stokes beam, respectively. (c) The Stokes beam intensity at $z = z_{max}$ is plotted with respect to $x/z_R$. The red open circle and blue line correspond to the pump beam intensities without and with the doughnut-shaped pump beam, respectively.

Using the same set of parameters and assuming that the doughnut-shaped pump beam's $|P_0^{l_p=1}(z_0)|$ is 0.752 MV (Table 1), I calculated the Stokes beam intensity distribution on the $x$-$y$ plane (Figure 7(a)). Due to the SRG of the Gaussian Stokes beam, the peak position shifts beyond the focal point, specifically at $z_{max} = 0.3z_R$. In Figure 7(b), the intensity of the Stokes beam along the optic axis is plotted. The manifestation of SRG is evident as the intensity of the Stokes beam increases at the focus but diminishes due to the beam divergence after passing the focal point.

Now, the case that the intensity of the doughnut-shaped pump beam is increased by a factor of ten, i.e., $|P_0^1(z_0)| = 0.752\sqrt{10} = 1.68$ MV is considered. As illustrated in Figure 8(b), the Stokes beam intensity increases, but its lateral width is not much different from the Stokes beam in the absence of the doughnut pump, as shown in Figure 8(c). In order to understand the underlying mechanism behind this observation, the values of $|S_n^0(z)|$ at the peak of $z_{max} = 4.3z_R$ are plotted in Figure 8(d), which are the expansion coefficients of the radial LG modes, $V_n^0(r, \phi, z)$, used to describe the Stoke beam. Despite the initial conditions $|S_{n=0}^0(z_0)| \neq 0$ and $S_{n>0}^0(z_0) = 0$, the SR-induced mode-mixing processes result in $S_{n>0}^0(z)$ becoming nonzero. Consequently, the contributions from $V_{n>0}^0(r, \phi, z)$ with one or more nodes to the Stokes beam profile become non-negligible. However, the resulting Stokes beam width on the lateral plane does not exhibit substantial changes.

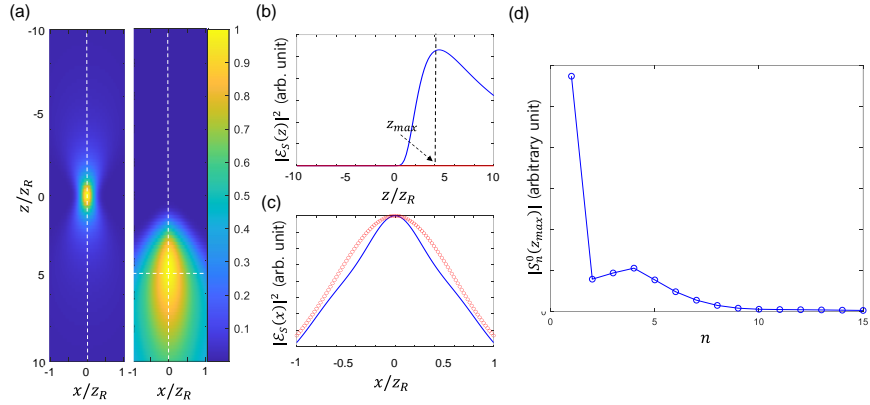

**Figure 8.** The intensity distributions of the Stokes beam in the absence and presence of a doughnut-shaped pump beam with an intensity approximately ten times stronger than that of the Stokes. (a) The intensity distributions of the Stokes beam on the $x$-$z$ plane without (left) and with (right) the doughnut-shaped pump beam. (b) The Stokes beam intensity at $x = 0$ and $y = 0$ is plotted with respect to $z/z_R$. The red and blue lines correspond to those without and with the doughnut-shaped pump beam, respectively. (c) The Stokes beam intensity distribution at $z = z_{max}$ is plotted with respect to $x/z_R$. The red open circle is the Stokes beam intensity without the pump beam, whereas the blue line corresponds to the Stokes beam intensity in the presence of the doughnut-shaped pump beam. (d) The absolute magnitudes of $|S_n^0(z_{max})|$ are plotted for different $n$'s.

The intensity of the Stokes beam along the optic axis exhibits a considerable increase by several orders of magnitude, as evident in Figure 8(b). However, this outcome is merely an artifact arising from the theoretical approximation in Eq. (52), which assumes that the pump beam intensity is sufficiently strong enough to be considered depletion-free. In the following section IV, I will address this limitation by numerically solving the coupled pump-Stokes equations. This numerical approach will illustrate that the SRG of the Stokes beam is constrained by the coupled pump beam intensity. Nevertheless, even with these computational results in section III, which involve a few approximations, the findings still indicate that the SRG of the Stokes beam does not lead to narrowing when the radial intensity distribution of the pump beam adopts one of the toroids with different central radii.

## 4. Finite difference calculations of pump-Stokes coupled equations

In the preceding section, I treated the nonlinear differential equations for the pump and Stokes beams independently, under the assumption that one of the two beams could be considered much stronger and, thus, independent of $z$. Consequently, the continuous increase in the Gaussian Stokes beam observed is an artifact stemming from the unrealistic assumption that the pump intensity is infinitely strong.

In this section, the coupled equations in Eqs. (25) and (26) are solved using a simple finite-difference method:

$$P_{n'}^{lp}(z + \delta z) = P_{n'}^{lp}(z) + \delta z \sum_n \Lambda_{n',n}^{lp}(z) P_n^{lp}(z) \tag{55}$$

$$S_{n'}^{ls}(z + \delta z) = S_{n'}^{ls}(z) + \delta z \sum_n \Gamma_{n',n}^{ls}(z) S_n^{ls}(z), \tag{56}$$

where the loss and gain matrix elements are

$$\Lambda_{n',n}^{lp}(z) = i \frac{2\pi k_p}{n_p^2} \chi_R^{(3)*} \int_0^{2\pi} d\phi \int_0^\infty dr\, r\, U_{n'}^{lp^*}(r,\phi,z) \left| \sum_p S_p^{ls}(z) V_p^{lp}(r,\phi,z) \right|^2 U_n^{lp}(r,\phi,z) \tag{57}$$

$$\Gamma_{n',n}^{ls}(z) = i \frac{2\pi k_S}{n_S^2} \chi_R^{(3)} \int_0^{2\pi} d\phi \int_0^\infty dr\, r\, V_{n'}^{ls^*}(r,\phi,z) \left| \sum_p P_p^{lp}(z) U_p^{lp}(r,\phi,z) \right|^2 V_n^{ls}(r,\phi,z). \tag{58}$$

For each finite-difference step $\delta z$ along the $z$-axis, it is necessary to compute the double integrations in Eqs. (57) and (58). Thus obtained values are then substituted into Eqs. (55) and (56) to derive $P_{n'}^{lp}(z + \delta z)$ and $S_{n'}^{ls}(z + \delta z)$. Repeating these calculations, iteratively, starting from the front surface of the sample at $z = -L$ (Figure 3) and progressing to the desired z-position within the Raman medium, yields the complete sets of expansion coefficients, $\left\{ P_n^{lp}(z) \right\}$ and $\left\{ S_n^{ls}(z) \right\}$. Utilizing these results and the LG functions for the pump and Stokes beams, $\left\{ U_n^{lp}(r,\phi,z) \right\}$ and $\left\{ V_n^{ls}(r,\phi,z) \right\}$, in Eq. (16), allows the determination of the amplitudes and phases of the two beams at any point in space.

### 4.1 Numerical simulation of SRG: Strong doughnut pump and weak Gaussian Stokes

In Figure 4 (Figure 6), I examined a scenario where the intensity of the doughnut-shaped Stokes (pump) beam was comparable to that of the Gaussian pump (Stokes) beam. In other words, the SRL and SRG of the pump and Stokes beams, respectively, were not exceptionally strong. In such instances, the evolution of the pump and Stokes beams was determined by solving Eqs. (55) and (56) with (57) and (58). Although not explicitly depicted here, the results closely resemble those presented in Figures 4 and 6. Essentially, the beam intensity profiles of the pump and Stokes beams exhibit minimal changes at the weak loss or gain limit. This observation suggests that the approximations employed to decouple the differential equations for the pump and Stokes beams were reasonable and acceptable.

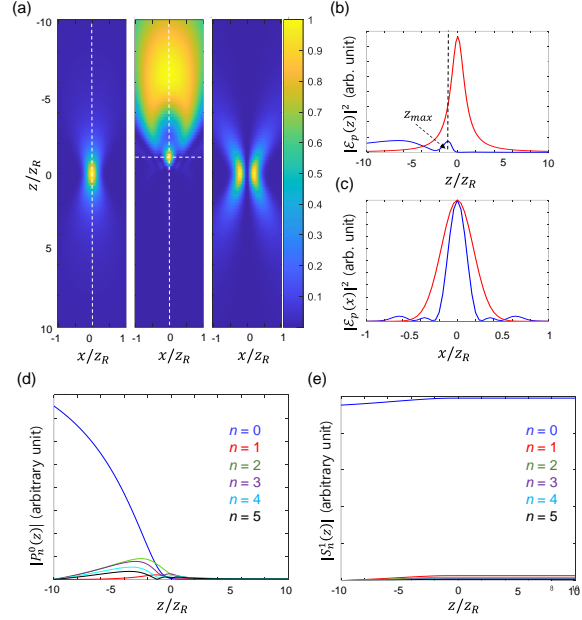

**Figure 9.** (a) The intensity distributions of the pump beam in the absence (left) and presence (middle) of a doughnut-shaped Stokes beam with an intensity approximately ten times stronger than that of the pump. The intensity distribution of the SRG Stokes beam is also shown in the right panel. (b) The pump beam intensity at $x = 0$ and $y = 0$ is plotted with respect to $z/z_R$. The red and blue lines correspond to those without and with the doughnut-shaped Stokes beam, respectively. (c) The pump beam intensity distribution at $z = z_{max}$ is plotted with respect to $x/z_R$. The red line is the pump beam intensity without the Stokes beam, whereas the blue line corresponds to the pump beam intensity in the presence of the doughnut-shaped Stokes beam. The absolute magnitudes of $|P_n^0(z)|$ and $|S_n^1(z)|$ are plotted in (d) and (e), respectively.

However, with the increase in the intensity of the doughnut-shaped Stokes beam compared to the pump, the loss factor increases, reading to the mixing of higher-order LG modes. Specifically, I examine the combination of the $LG_{00}$ pump with an initial amplitude of $|P_0^0(z_0)| = 0.752$ MV and the $LG_{01}$ Stokes beam with an amplitude of $|S_0^1(z_0)| = 0.752\sqrt{10} = 1.68$ MV. In Figure 9, the pump beam intensity distribution on the $x$-$z$ plane is displayed. For comparisons, the intensity distributions of the pump $LG_{00}$ mode in the absence of a Stokes beam and that of the doughnut Stokes beam are plotted in Figure 9(a) (left and middle panels). Since the Stokes beam intensity is stronger by a factor of 10 than that of the pump beam, the Stokes beam profile does not change much. However, the pump beam intensity (the blue line in Figure 9(b)) experiences a substantial reduction due to the SRL process induced by the $LG_{01}$ Stokes beam. This $z$-dependent SRL-pump beam intensity should be contrasted with the blue line in Figure 5(b). Although the intensity profiles along the $z$-axis in Figures 5(b) and 8(b) are approximately similar, their detailed $z$-dependences appear different. This observation can be attributed to the fact that the intrinsically coupled SRG process of the Stokes beam was not considered when obtaining the numerical calculation results, assuming that the Stokes beam intensity remains constant throughout the Raman medium.

At the peak z-position at $z = z_{max}$ (Figure 9(b)), I obtained the pump beam intensity distribution (blue line) along the $x$-axis in Figure 9(c), which is then compared with the pump $LG_{00}$ mode (red open circle) in the absence of the Stokes beam. Due to the SRL of the pump beam in the region where it overlaps with the doughnut-shaped Stokes beam, the net effect narrows the lateral pump beam intensity distribution, revealing notable side lobes. To understand this change in the three-dimensional pump beam intensity distribution, I examined the z-dependent LG mode expansion coefficients $\{P_n^0(z)\}$ and $\{S_n^1(z)\}$ (see Figures 9(d) and 9(e)).

Firstly, the initial Stokes beam in the $LG_{01}$ mode gains intensity slightly (blue line in Figure 9(e)). Secondly, the major component, $LG_{00}$ mode in the pump, completely loses its intensity at $z > 0$. Interestingly, the other radial LG modes, $LG_{n0}$ ($n \geq 1$), become mixed in, forming part of a superposition state of the SRL pump beam. This result is consistent with the observation of mode mixing depicted in Figure 5(d).

## 4.2 Numerical simulation of SRG: Strong doughnut pump and weak Gaussian Stokes

When the intensity of the doughnut pump beam is ten times stronger than that of the Gaussian Stokes beam, both the SRG of the Stokes beam and the SRL of the doughnut pump beam occur simultaneously. To quantitatively understand these coupled processes, I carried out finite-difference calculations of Eqs. (55) and (56) with $|P_0^1(z_0)| = 1.68$ MV and the $LG_{00}$ Stokes beam with $|S_0^0(z_0)| = 0.752$ MV.

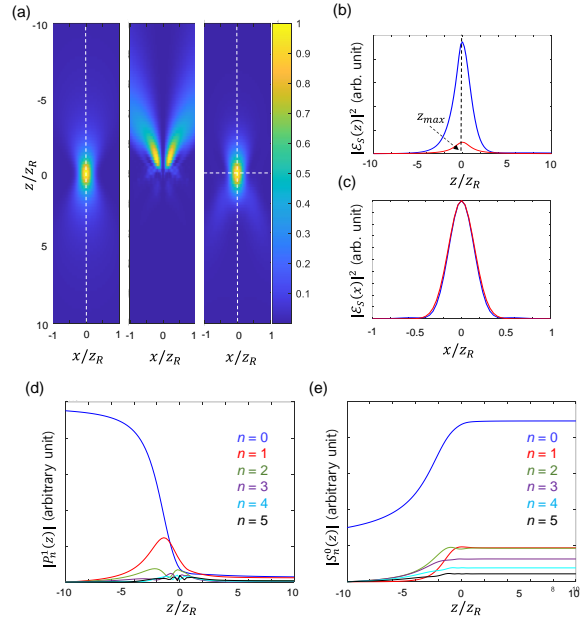

**Figure 10.** (a) The intensity distributions of the Gaussian Stokes beam in the absence (left) and presence (right) of a doughnut-shaped pump beam with an intensity approximately ten times stronger than that of the Stokes. The intensity distribution of the SRL pump beam is also shown in the middle panel. (b) The Stokes beam intensity at $x = 0$ and $y = 0$ is plotted with respect to $z/z_R$. The red and blue lines correspond to those without and with the doughnut-shaped pump beam, respectively. (c) The Stokes beam intensity distribution at $z = z_{max}$ is plotted with respect to $x/z_R$. The red line is the Stokes beam intensity without the pump beam, whereas the blue line corresponds to the Stokes beam intensity in the presence of the doughnut-shaped pump beam. The absolute magnitudes of $|P_n^1(z)|$ and $|S_n^0(z)|$ are plotted in (d) and (e), respectively.

The intensity distributions of the SRL-pump and SRG-Stokes beams are shown in Figure 10(a) (middle and right panels). Additionally, the Stokes $LG_{00}$ mode in the absence of the pump beam is presented for comparison in Figure 10(a) (left panel). Despite the SRS process, the profile of the Stokes beam, except for its gain in intensity, remains largely unaffected by the presence of the doughnut-shaped pump beam (Figures 10(b) and 10(d) for the Stokes beam intensity distributions along the $z$- and $x$-axes, respectively).

In the SRL of the pump, it is evident that the initial $LG_{01}$ mode expansion coefficient (blue line in Figure 10(d)) decreases as it propagates through the Raman medium. However, the expansion coefficients of higher-order $LG_{0l}$ modes (where $l > 1$) increase and then decrease to zero (Figure 10(d)). Also, the Stokes beam transforms into a superposition of radial $LG_{n0}$ modes (where $n \geq 1$) (Figure 10(e)). Nevertheless, the doughnut shape of the pump beam does not induce substantial changes in the 3D intensity distribution of the initial Gaussian Stokes beam, as demonstrated here (right panel in Figure 10(a)).

### 4.3 Numerical simulation of SRL: Weak Gaussian pump and strong high rotational-index Stokes

In subsection IV-A above, I assumed that the pump beam is weak and Gaussian, while the Stokes beam is strong and in an $LG_{01}$ mode.[19] However, as the rotational index $l$ of the $LG_{0l}$ Stokes beam increases, the radius of the central hole of a given doughnut-shaped Stokes beam increases (Figure 11). Consequently, the mode mixing effects on the transmitted pump beam, which are dictated by the overlap integrals determining the loss and gain matrices in Eqs. (57) and (58), would vary for different orbital angular momentum indices $l$ of the injected Stokes beam.

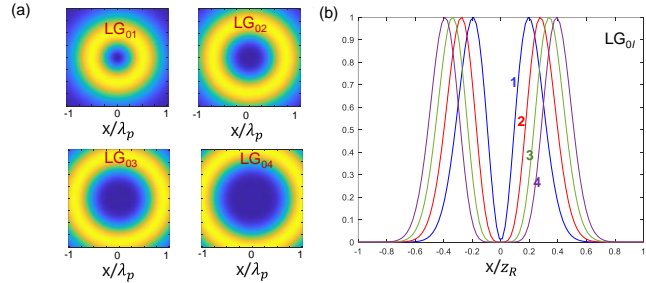

**Figure 11.** (a) The intensity distributions of doughnut-shaped Stokes beam in one of the $LG_{0l}$ modes for $l$ = 1-4. (b) The Stokes intensity distributions are plotted with respect to $x/z_R$.

In Figure 12(a), the resulting pump beam intensity distributions on the $x$-$z$ plane are presented. The first panel in Figure 12(a) displays the pump $LG_{00}$ mode in the absence of the Stokes beam. As the ring radius of the doughnut-shaped Stokes beam increases, the pump beam intensity distribution undergoes a notable change in axial resolution along the $z$-axis, even though the lateral resolution on the $x$-$y$ plane remains relatively constant. In order to showcase the enhanced axial resolution, the pump beam intensity on the optic axis is plotted with respect to $z$ in Figures 12(b) and 12(c), where the Stokes beam is in $LG_{01}$ and $LG_{04}$ modes, respectively. Interestingly, due to the variation in the Gouy phase of $LG_{n0}$ for different radial indices $n$, the superposition of pump $LG_{n0}$ modes exhibits destructive interferences, resulting in the narrowing of the pump beam intensity distribution along the $z$-axis.

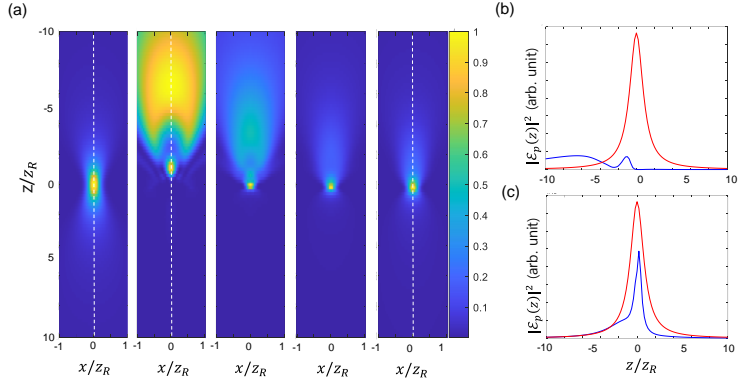

**Figure 12.** (a) The intensity distributions of the pump beam in the absence (1st panel) and presence of a doughnut-shaped $LG_{0l}$ Stokes beam with an intensity approximately ten times stronger than that of the pump. The intensity distribution in the ($l$+1)th panel is the case that the Stokes beam is in the $LG_{0l}$ mode. (b) The pump beam intensity at $x = 0$ and $y = 0$ is plotted with respect to $z/z_R$. The red and blue lines correspond to those without and with the doughnut-shaped $LG_{01}$ Stokes beam, respectively. (c) The pump beam intensity at $x = 0$ and $y = 0$ is plotted with respect to $z/z_R$. The red and blue lines correspond to those without and with the doughnut-shaped $LG_{04}$ Stokes beam, respectively.

To understand this phenomenon of SR-induced narrowing of the pump beam, the absolute values, real parts, and imaginary parts of $P_n^0(z)$ for $n = 0$-6 are plotted in Figure 13. The sign of the $Re[P_1^0(z)]$, associated with the $LG_{10}$ mode, is negative and opposite to that of the $Re[P_0^0(z)]$ associated with the fundamental Gaussian mode (Figure 13(d)).

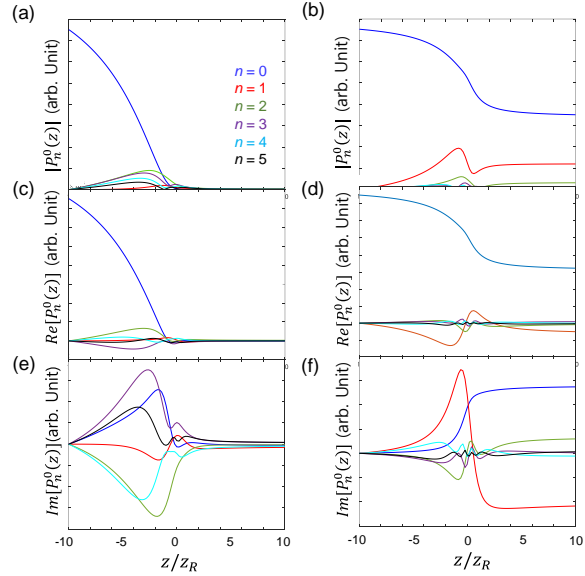

**Figure 13.** The absolute magnitude, real part, and imaginary part of $P_n^0(z)$ are plotted in (a), (c), and (e), when the doughnut-shaped Stokes beam is in the $LG_{01}$ mode. Those of $P_n^0(z)$ are plotted in (b), (d), and (f), when the doughnut-shaped Stokes beam is in the $LG_{04}$ mode.

In the case where the Stokes beam is in an $LG_{01}$ mode, the magnitude of $P_1^0(z)$ remains comparatively small (Figure 13(a)). However, as the rotational index of the Stokes beam increases from $l = 1$ to $l = 4$, i.e., with an increasing radius of the doughnut ring, the magnitude of negative $Re[P_1^0(z)]$ increases as $z$ approaches the focal point (Figure 13(d)). Consequently, the two modes, $LG_{00}$ and $LG_{10}$, strongly interfere with each other destructively. Similarly, the phase change of the imaginary part value, $Im[P_1^0(z)]$, is notable around the focal position at $z = 0$ (Figure 13(f)). Analyzing the real and imaginary parts of $P_n^0(z)$ allows for deducing the phase angles of $P_n^0(z)$. It turns out that there is a $\pi$ phase-difference between $P_0^0(z)$ and $P_1^0(z)$ at the focus ($z = 0$), leading to destructive interference that narrows the pump beam intensity distribution.

## 5. Summary

In this paper, I presented theoretical descriptions and numerical calculation results for stimulated Raman-induced beam focusing when the conjugate beam carries nonzero orbital angular momentum. I considered two distinct SRS schemes, one involving a pair of a weak LG00-pump and a strong LG0l-Stokes beam, and the other involving a strong LG01-pump and a weak LG00-Stokes beam. In these schemes, the SRL and SRG processes are measured, respectively.

I derived coupled equations for the expansion coefficients by expanding the pump and Stokes beams with higher-order Laguerre-Gauss modes. When one of the two beam intensities is sufficiently large to be considered constant, i.e., depletion-free pump or gainless Stokes beam, the corresponding differential equations for the pump and Stokes beams along the propagation direction ($z$-axis) become decoupled. I obtained an analytical solution for the resulting nonlinear differential equation.

Using a finite-difference method, I solved the coupled differential equations for the LG mode-expansion coefficients associated with the pump and Stokes beams. The numerical results show that the intensity distribution of the $LG_{00}$ pump along the $z$-axis becomes narrower than that without an SRS process induced by a doughnut-shaped Stokes beam. Remarkably, this pump beam focusing is more pronounced as the rotational index or the orbital angular momentum of the Stokes beam increases.

The theoretical findings presented in this paper offer valuable insights for demonstrating coherent Raman imaging with enhanced axial spatial resolutions and for optical detection of the orbital angular momentum (OAM) components of light, which are particularly relevant in OAM light communications.[30, 37]

**Funding**

This work was supported by the Institute for Basic Science (IBS-R023-D1).